\documentclass[twocolumn,jcp,aip,amscd]{revtex4}
\usepackage{amssymb}
\usepackage{amsmath}
\usepackage{amsthm}
\usepackage{graphicx}
\usepackage{color}

\begin{document}

\title{Actinide chemistry using singlet-paired coupled 
cluster and its combinations with density functionals}
\author{Alejandro J. Garza}
\author{Ana G. Sousa Alencar}
\affiliation{Department of Chemistry, Rice University, Houston, Texas, 77251-1892, USA}
\author{Gustavo E. Scuseria}
\affiliation{Department of Chemistry and Department of Physics and Astronomy, Rice
University, Houston, Texas, 77251-1892, USA}
\date{\today}

\begin{abstract}
  Singlet-paired coupled cluster doubles (CCD0) is 
  a simplification of CCD that relinquishes 
  a fraction of dynamic correlation in order to 
  be able to describe static correlation. 
  Combinations of CCD0 with density functionals that recover 
  specifically the dynamic correlation missing in the 
  former have also been developed recently.
  Here, we assess the accuracy of 
  CCD0 and CCD0+DFT (and variants of these using 
  Brueckner orbitals)
  as compared to 
  well-established quantum chemical methods 
  for describing ground-state properties of singlet 
  actinide molecules. 
  The $f^0$ actinyl series 
  (UO$_2^{2+}$, NpO$_2^{2+}$, PuO$_2^{2+}$),
  the isoelectronic NUN, and 
  Thorium (ThO, ThO$^{2+}$) and Nobelium 
  (NoO, NoO$_2$) oxides are studied.
  
\end{abstract}

\pacs{0000-1111-222}
\maketitle

\label{I1}

\section{Introduction}

Actinide chemistry represents a challenge for
experimental approaches due to the high 
toxicity and radioactivity of 
actinide compounds.
Accurate computational models are therefore 
particularly valuable in this area of chemistry.
An example of this was the theoretical prediction of 
NUO$^+$~\cite{Pyyko1994}
 and its subsequent discovery by mass spectroscopy~\cite{Heinemann1995}.
However, actinide chemistry is also 
challenging for common quantum chemical 
approximations: The presence of multiple 
degenerate, partially filled $f$ orbitals
leads to substantial static correlation. 
Typical techniques for handling static correlation 
have severe limitations such as lack of size-consistency 
and size-extensivity and, most restrictively, 
a combinatorial increase in computational cost 
with system size~\cite{Garza2015b}. 
Furthermore, many of these approaches 
may miss important dynamic correlation. 
Other popular methods of quantum chemistry 
such as Kohn--Sham density functional theory 
(KS-DFT) or single-reference coupled cluster (CC) are unreliable 
when static correlation is present~\cite{Bulik2015} 
(common CC methods may even diverge or yield complex 
correlation energies).
Recent advances in computational actinide 
chemistry have been reviewed in Refs.~\cite{Kovacs2015,Wang2012}

Recently, techniques that modify the cluster operator 
of CC doubles (CCD) in order to describe static 
correlation---at the cost of neglecting some 
dynamic correlation---have been proposed in the literature. 
These include 
pair CCD~\cite{Limacher2013,Limacher2014,Tecmer2014,Boguslawski2014,%
Stein2014,Henderson2014}
 (pCCD), as well as singlet-paired 
CCD and Brueckner doubles~\cite{Bulik2015,Garza2015} (CCD0 and BD0, respectively). 
All of these methods 
have polynomial scaling and 
are size-consistent and size-extensive
(provided that the reference determinant be size-consistent and 
size-extensive, which may demand symmetry breaking~\cite{Hoyos2011}). 
Furthermore, approaches to incorporate the dynamic correlation 
absent in CCD0 and BD0 via density functionals (CCD0+DFT)
have been developed~\cite{Garza2015}. 
A recent study~\cite{Tecmer2015} using pCCD suggests 
this new type of CC \textit{ans\"{a}tze} to be promising 
for applications in actinide chemistry. 
Here, we assess the accuracy of 
CCD0 and CCD0+DFT (and their BD0 variants) as compared to 
well-established quantum chemical methods 
for describing ground-state properties of singlet 
actinide molecules. 
The $f^0$ actinyl series 
(UO$_2^{2+}$, NpO$_2^{2+}$, PuO$_2^{2+}$),
the isoelectronic NUN, and 
Thorium (ThO, ThO$^{2+}$) and Nobelium 
(NoO, NoO$_2$) oxides are studied.

\section{Theory and Methods}

\subsection{Singlet-Paired Coupled cluster Doubles (CCD0)}

We give here a minimal description of CCD0; for further details 
see Ref.~\cite{Bulik2015}.
Like standard CC methods, CCD0
uses an exponential wavefunction~\cite{Bulik2015}
\begin{equation}
  | \Psi_\text{CCD0} \rangle = e^{T_2^{[0]}} | \Phi_\text{RHF} \rangle,
  \label{eq:ansatz}
\end{equation}
where $| \Phi_\text{RHF} \rangle$ is a
restricted Hartree--Fock determinant and 
$T_2^{[0]}$ is the cluster operator 
of singlet-paired double excitations:
\begin{equation}
  T_2^{[0]} = \frac{1}{2} \sum \limits_{ijab}
  \sigma_{ij}^{ab} P_{ab}^\dag P_{ij},
  \label{eq:T2}
\end{equation}
where $ij$ and $ab$ are indices for occupied and virtual orbitals, 
respectively; 
the amplitudes obey the relation 
$\sigma_{ij}^{ab} = (t^{a\uparrow b\downarrow}_{i\uparrow j\downarrow} + 
    t^{b\uparrow a\downarrow}_{i\uparrow j\downarrow})/2$;  
 and
\begin{equation}
  P_{ij}  = \frac{1}{\sqrt{2}}  \left( c_{j\uparrow} 
  c_{i\downarrow} +   c_{i\uparrow}  c_{j\downarrow} \right).  
    \label{eq:Pab} 
\end{equation}
That is, $P_{ab}$ acting on $| \Phi_\text{RHF} \rangle$
gives a singlet and $e^{T_2^{[0]}}$ 
contains contributions from all singlet-paired excitations. 
In standard CCD, the cluster operator contains singlet- and 
triplet-paired contributions, hence capturing more correlation. 
However, it is the combination of the singlet- and
triplet-paired components that causes the failure
of CCD (and CCSD, CCSDT, \textit{etc.})
in strongly correlated 
systems, which results in unphysical correlation 
energies~\cite{Bulik2015} (\textit{i.e.}, too large, divergent, 
or complex energies).
Thus, CCD0 relinquishes a fraction of the correlation 
in exchange for safeguard against this breakdown.

Singlet-paired Brueckner doubles (BD0) is analogous to CCD0,
the only difference being that 
approximate Brueckner~\cite{Dykstra1997,Handy1989,Kobayashi1991}, 
rather than RHF, orbitals are used as reference:
\begin{equation}
  | \Psi_\text{BD0} \rangle = e^{T_2^{[0]}} | \Phi_\text{BD} \rangle.
  \label{eq:bd0}
\end{equation}
Specifically, 
the approximate Brueckner orbitals in $| \Phi_\text{BD} \rangle$
are those
which zero out the amplitudes of single substitutions in a 
model subspace of single and double substitutions (just like 
in standard Brueckner doubles).

\subsection{Combination with Density Functionals (CCD0+DFT)}

The different flavors of CCD0+DFT are discussed in 
detail in Ref.~\cite{Garza2015}; we just provide here 
a brief explanation of these techniques for the sake 
of clarity. 
There are two categories of CCD0+DFT methods:
one that adds parallel spin correlation to CCD0, and 
another that adds the full contributions from triplet-paired 
excitations.
The first one is derived by noting that the
CCD0 correlation energy is
\begin{equation}
  E_c^\text{CCD0} = \sum \limits_{ijab}
  \sigma_{ij}^{ab} v^{a\uparrow b\downarrow}_{i\uparrow j\downarrow}, 
  \label{eq:Ec1}
\end{equation}
where $v^{a b}_{i j} = \langle i j |  a b \rangle$
is a two-electron integral in the Dirac notation.
Hence, there are no contributions to the correlation from pairs of
same-spin electrons. 
One can thus add (without double counting) the equal spin correlation to 
CCD0 using a density functional approximation (DFA). 
For a singlet, the correlation energy would be
\begin{equation}
  E_c^\text{CCD0+pDFT} = E_c^\text{CCD0} + 
  2  E_{c \, \uparrow \uparrow}^\text{DFA} [n_\uparrow, n_\downarrow], 
  \label{eq:Ec2}
\end{equation}
where the ``p'' in pDFT is for ``\textit{parallel-spin}'' and 
$E_{c \, \alpha \alpha}^\text{DFA}$ is the DFA correlation for the 
spin-$\alpha$ density.

The second category of CCD0+DFT is derived by noting that 
the full double excitations cluster operator 
$T_2$ used in the latter can be expressed as
\begin{equation}
  T_2 = T_2^{[0]} + T_2^{[1]}, 
\end{equation}
where $T_2^{[0]}$ is defined above and 
$T_2^{[1]}$ is the triplet-paired component of $T_2$
\begin{equation}
  T_2^{[1]} = \frac{1}{2} \sum \limits_{ijab}
  \pi_{ij}^{ab} \mathbf{Q_{ab}^\dag} \cdot 
  \mathbf{Q_{ij}}
  \label{eq:T21}
\end{equation}
where $\mathbf{Q_{ij}}$ is a vector 
$\mathbf{Q_{ij}} = ( Q_{ij}^{+}, Q_{ij}^{0}, Q_{ij}^{-} )^t$
whose components are
\begin{equation}
  Q_{ij}^{+} =   c_{j\uparrow}  c_{i\uparrow} , \quad 
  Q_{ij}^{-} =   c_{j\downarrow}  c_{i\downarrow} ,
  \label{eq:Q1}
\end{equation}

\begin{align}
  Q_{ij}^0 & = \frac{1}{\sqrt{2}}  \left( c_{j\uparrow} 
  c_{i\downarrow} -   c_{i\uparrow}  c_{j\downarrow} \right) 
  \label{eq:Q0} \\
  & = \frac{1}{\sqrt{2}}  \left( c_{j\uparrow} 
  c_{i\downarrow} +   c_{j\downarrow} c_{i\uparrow} \right). \nonumber
\end{align}
Thus, CCD0 misses not only parallel spin correlation, but also 
the $m = 0$ channel of $T_2^{[1]}$. 
For a closed shell, the $m = +1, 0,$ and $-1$ components 
of of $T_2^{[1]}$ contribute equally to the correlation. 
We may therefore incorporate the opposite spin correlation 
missing in CCD0+pDFT by adding $E_c^\text{DFA} [n_\uparrow, 0]$ once more to
the (closed-shell) energy 
\begin{equation}
  E_c^\text{CCD0+tDFT} = E_c^\text{CCD0} + 
  3   E_{c \, \uparrow \uparrow}^\text{DFA} [n_\uparrow, n_\downarrow],
  \label{eq:Etot2}
\end{equation} 
where the ``t'' in tDFT indicates that the full contributions from
the \textit{triplet-paired} component of $T_2$ are being 
taken into account. 

\subsection{Parallel Spin Functionals for CCD0+DFT}

The CCD0+DFT methods described above 
require a spin resolution for DFA correlation in order 
to approximate
$E_{c \, \uparrow \uparrow}^\text{DFA} [n_\uparrow, n_\downarrow]$.
For completeness, we discuss this topic briefly here. 
The simplest spin resolution for the DFA correlation is
the exchange-like \textit{ansatz} 
of Stoll \textit{et al.}~\cite{Stoll1978}
\begin{equation}
  E_{c \,\uparrow \uparrow}^\text{DFA} [n_\uparrow, n_\downarrow] =
  E_c^\text{DFA} [n_\uparrow, 0]. 
  \label{eq:Stoll}
\end{equation} 
This equation can be used for the local density approximation 
(LDA), generalized gradient approximations (GGAs), and 
meta-GGAs that are rooted on the homogeneous electron gas.
(Not all DFAs have a meaningful spin resolution. 
The Lee--Yang--Parr~\cite{Lee1988}
 functional, for example, models all correlation 
as being opposite spin.)
The use of meta-GGAs is most desirable because these functionals 
are free of one-electron self-interaction, and thus BD0+DFT 
with a meta-GGA is exact for two-electron singlets
using the spin resolution of Eq.~\ref{eq:Stoll}.
Here, we use two nonempirical meta-GGAs in combination 
with CCD0 and BD0:
the Tao--Perdew--Staroverov--Scuseria~\cite{Tao2003}
(TPSS) functional, 
and the strongly constrained and appropriately normed (SCAN) 
functional of Sun \textit{et al.}~\cite{Sun2015}.

Equation~\ref{eq:Stoll} is an educated guess which 
is exact only for fully spin-polarized densities and 
in the high-density limit of the uniform electron gas~\cite{Gori-Giorgi2004}.
In the case of SCAN, however, it is possible to compose 
an improved guess for the same-spin correlation~\cite{Garza2015}.
This is because SCAN constructs the correlation energy density 
$\varepsilon_c$ as~\cite{Sun2015} 
\begin{equation}
  \varepsilon_c = 
  \varepsilon_c^1  + 
  f_c(\alpha) \left[ \varepsilon_c^{0}  - 
    \varepsilon_c^{1}  \right], 
\end{equation}
where $f_c(\alpha)$ is a function that depends on the 
kinetic energy density
(see Supporting Information of Ref.~\cite{Sun2015}), 
and $\varepsilon_c^{\alpha = 0}$ and 
$\varepsilon_c^{\alpha = 1}$ are the single orbital and 
uniform density limits, respectively, for the 
correlation energy density.
It is thus logical to write the spin-up 
correlation energy density for SCAN as 
\begin{equation}
  \varepsilon_c^{\uparrow \uparrow} = 
  \varepsilon_c^{1 \, \uparrow \uparrow}  + 
  f_c(\alpha) \left[ \varepsilon_c^{0 \, \uparrow \uparrow}  - 
    \varepsilon_c^{1 \, \uparrow \uparrow}  \right]. 
\end{equation}
Furthermore, 
$\varepsilon_c^{0 \, \uparrow \uparrow} = 0$
because there is no parallel-spin correlation for two electrons in the 
same spatial orbital.
Hence, 
\begin{equation}
  \varepsilon_c^{\uparrow \uparrow} = 
  \varepsilon_c^{1 \, \uparrow \uparrow}  -
  f_c(\alpha) 
    \varepsilon_c^{1 \, \uparrow \uparrow}
\end{equation}
so that $\varepsilon_c^{\uparrow \uparrow}$  depends only on the 
uniform density limit of the spin-up correlation energy density, 
$\varepsilon_c^{1 \, \uparrow \uparrow}$.
The spin resolution of $\varepsilon_c$ in the uniform 
electron gas has been parametrized by 
Gori-Giorgi and Perdew~\cite{Gori-Giorgi2004} in terms 
for fractions fractions 
$F_{\sigma \sigma'}$ such that 
$\varepsilon_c^{\sigma \sigma'} = \varepsilon_c F_{\sigma \sigma'}$.
Thus, we compute
$\varepsilon_c^{1 \, \uparrow \uparrow} = 
\varepsilon_c  F_{\uparrow \uparrow} $ using
the expression for $F_{\uparrow \uparrow}$ given in 
Equation 9 of Ref.~\cite{Gori-Giorgi2004}.
We term the variation of SCAN using this spin \textit{resolution}
as ``rSCAN''.
For convenience of the reader, Table~\ref{tab:methods} summarizes the 
CCD0+DFT methods described and employed in this work.
Further details regarding CCD0+DFT, including discussion 
about the spin resolutions, may be consulted in 
Ref.~\cite{Garza2015}.

\renewcommand{\arraystretch}{1.35}
\begin{table}
  \begin{center}
    \caption{Summary of CC0+DFT methods employed here. 
      The notation and closed-shell energy formulas are given;
      CC0 can refer to CCD0 or BD0 and the densities are 
      from the RHF or Brueckner reference determinants, respectively; 
      the ``p'' in pDFT is for \textit{parallel spin}; the ``t'' in tDFT 
      is for \textit{triplet-pairing component};
      DFT may refer to TPSS or SCAN; and
      $E_{c \, \uparrow \uparrow}^\text{rSCAN}$ is the spin-up 
      SCAN correlation using the spin
      \textit{resolution} of Section II.C.}
    \label{tab:methods}
    \scalebox{0.95}{
      \begin{tabular*}{0.5\textwidth}{@{\extracolsep{\fill}}  ll  }
        \hline
        Method & Energy Formula  \\
        \hline 
        CC0+pDFT & $ E^\text{CC0} + 
        2 E_c^\text{DFA}[n_\uparrow,0]$ \\
        CC0+tDFT & $ E^\text{CC0} + 
        3 E_c^\text{DFA}[n_\uparrow,0]$ \\
        CC0+prSCAN & $ E^\text{CC0} + 
        2 E_{c \, \uparrow \uparrow}^\text{rSCAN} [n_\uparrow, n_\downarrow ]  $\\
        CC0+trSCAN & $ E^\text{CC0} + 
        3 E_{c \, \uparrow \uparrow}^\text{rSCAN} [n_\uparrow, n_\downarrow ]  $ \\
 \hline    
      \end{tabular*}
    }
  \end{center}
\end{table}
\renewcommand{\arraystretch}{1.0}

\subsection{Computational Details} 

All calculations were carried out using a development 
version of \textsc{Gaussian}~\cite{Gaussian} in which the CCD0 and 
CCD0+DFT methods have been implemented.
As in Ref.~\cite{Garza2015}, 
CCD0+DFT calculations are done in a non-self-consistent 
manner: the DFA correlation is evaluated in a single-shot, 
post-CCD0 calculation with the densities from the 
reference determinant. 
CCD0+DFT geometry optimizations and 
harmonic vibrational frequencies
were computed numerically using a convergence threshold 
of $1 \times 10^{-9}$ Hartrees on the CCD0 energy 
and the largest of the preset grids in \textsc{Gaussian} for integrating
the density functional (\texttt{Integral=SuperFine} keyword).
CCD0 geometries and frequencies were evaluated 
analytically, and we verified that the numerical 
procedure for determining these properties with CCD0+DFT agreed with 
analytical results for CCD0.
These same specifications were followed in 
BD0 and BD0+DFT calculations. 
Unless otherwise indicated, all calculations employ 
the Stuttgart relativistic small-core effective core 
potential~\cite{Bergner1993,Kaupp1991,Dolg1993}
(RSC ECP) basis for actinide atoms, and the
aug-cc-pVDZ basis for the light atoms. These basis sets have 
been shown to be adequate for the type of calculations 
carried out here~\cite{Han2000}.
Spin-orbit coupling effects are neglected as they are 
not important for the closed-shell species studied 
in this work~\cite{Kovacs2015,Fromager2009}. 
The results reported here are all in 
gas phase media.

\section{Results and Discussion}

\subsection{Uranyl Cation (UO$_\mathbf{2}^{\mathbf{2+}}$)}

The uranyl ion UO$_2^{2+}$ is considered the most 
important of the actinyls: 
Nuclear reactors usually rely on uranium to fuel nuclear 
chain reactions, and 
UO$_2^{2+}$ is the most common form of uranium in 
aqueous solution. 
UO$_2^{2+}$ is highly toxic and its study is motivated 
by the need for knowledge regarding soluble 
actinide complexes, which are important for nuclear waste disposal 
and environmental transport~\cite{Kovacs2015}.
The uranyl cation has therefore been studied 
extensively~\cite{Kovacs2015,Han2000,Fromager2009,Jackson2008,%
Straka2001,Gagliardi2000}.
Here, we study this species in the gas phase due to the availability 
of data from high-level calculations to compare with, as the 
accuracy of CCD0 and CCD0+DFT for actinide compounds 
has not yet been established. 
To the best of out knowledge, there is no experimental data 
for the properties of UO$_2^{2+}$ here calculated.

\begin{table}
  \begin{center}
    \caption{Bond lengths ($R_e$ in \AA) and harmonic vibrational 
    frequencies ($\omega_e$ in cm$^{-1}$)
    for UO$_2^{2+}$ ($D_{\infty h}$) calculated 
    by various methods. Results computed in this work appear in the top 
    part of the table; results compiled from the literature at the 
    bottom. To the best of our knowledge, 
    no experimental data is available for this species. }
    \label{tab:uo2}
    \scalebox{0.90}{
      \begin{tabular*}{0.5\textwidth}{@{\extracolsep{\textwidth minus\textwidth}}  lcccc  }
        \hline
        This Work &    &                      &                     &                  \\
        Method & $R_e$ & $\omega_{\text{as}}$ & $\omega_{\text{s}}$ & $\omega_{\beta}$ \\
        \hline 
        CCD0        & 1.698 & 1129 & 1053 & 541 \\
        CCD0+pTPSS  & 1.681 & 1162 & 1091 & 508 \\
        CCD0+pSCAN  & 1.686 & 1143 & 1073 & 230 \\
        CCD0+prSCAN & 1.687 & 1148 & 1078 & 223 \\
        CCD0+trSCAN & 1.683 & 1157 & 1086 & 226 \\
        BD0         & 1.700 & 1143 & 1049 & 253 \\
        BD0+pTPSS   & 1.685 & 1130 & 1061 & 379 \\
        BD0+pSCAN   & 1.694 & 1112 & 1044 & 226 \\
        BD0+prSCAN  & 1.693 & 1121 & 1052 & 220 \\
        BD0+trSCAN  & 1.689 & 1126 & 1057 & 381 \\
        CCSD        & 1.696 & 1151 & 1059 & 202 \\
        CCSD(T)     & 1.702 & 1113 & 1025 & 192 \\
        HF          & 1.648 & 1293 & 1220 & 267 \\
        PBE         & 1.715 & 1086 & 985  & 123 \\        
        PBEh        & 1.684 & 1175 & 1082 & 187 \\
        LC-$\omega$PBE & 1.674 & 1213 & 1125 & 192 \\
        \hline 
        Literature &    &                      &                     &                  \\
        Method & $R_e$ & $\omega_{\text{as}}$ & $\omega_{\text{s}}$ & $\omega_{\beta}$ \\
        \hline 
        pCCD$^a$    & 1.669 & ---  & 1060 & --- \\
        CCSD(T)$^b$ & 1.690 & 1120 & 1035 & 178 \\
        MP2$^a$     & 1.745 & ---  & 854  & --- \\
        MP2$^c$     & 1.724 & 1052 & 941  & 277 \\
        CASSCF(10,10)$^a$ & 1.694 & --- & 1085 & --- \\
        CASSCF(12,12)$^a$ & 1.707 & --- & 1034 & --- \\
        CAS-srPBE(12,12)$^d$ & 1.684 & --- & ---- & --- \\
        CAS-srLDA(12,12)$^d$ & 1.684 & --- & ---- & --- \\
        CASPT2(12,12)$^e$  & 1.714 & 1153 & 1043 & --- \\ 
        CASPT2(12,12)$^f$  & 1.705 & 1066 & 959 & --- \\ 
 \hline    
 \multicolumn{5}{@{}l}{\footnotesize{$^a$From Ref.~\cite{Tecmer2015} using 
     the cc-pVDZ basis on O. }} \\
 \multicolumn{5}{@{}l}{\footnotesize{$^b$From Ref.~\cite{Jackson2008};
     RSC+3$g$ on U and aug-cc-pVQZ on O.}} \\
 \multicolumn{5}{@{}l}{\footnotesize{$^c$From Ref.~\cite{Straka2001}; 
     RSC+2$g$ on U and aug-cc-pVDZ on O. }} \\
 \multicolumn{5}{@{}l}{\footnotesize{$^d$From Ref.~\cite{Fromager2009}; 
     RECP ($14s13p10d8f6g$)/[$6s6p5d4f3g$]
     on U and }} \\
 \multicolumn{5}{@{}l}{\footnotesize{($4s5p1d$)/[$2s3p1d$] on O. }} \\
 \multicolumn{5}{@{}l}{\footnotesize{$^e$From Ref.~\cite{Gagliardi2000}; 
     RSC on U and $4s3p2d$ ANO-S on O. }} \\
 \multicolumn{5}{@{}l}{\footnotesize{$^f$From Ref.~\cite{Gagliardi2000}; 
     RSC on U and $4s3p2d1f$ ANO-L on O. }} \\
      \end{tabular*}
    }
  \end{center}
\end{table}

Table~\ref{tab:uo2} compiles predictions by various methods 
for the bond lengths ($R_e$) 
and harmonic vibrational frequencies ($\omega_e$)
of UO$_2^{2+}$. 
Some of these data have been taken from the literature 
and not all calculations use the same basis set; 
however, the results should be
roughly comparable because the bases are all of similar, good
quality.
The highest level methods in this Table are CCSD(T) and CASPT2(12,12); 
CCSD(T) results are considered to be reliable for 
uranyl~\cite{Kovacs2015,Straka2001,Jackson2008}. 
CCD0 is in good agreement with CCSD(T) except for a large 
overestimation of the bending frequency $\omega_\beta$. 
This problem persists in CCD0+pTPSS, but is 
alleviated by combinations of CCD0 with SCAN. 
Compared to CCD0,
BD0 provides a much better estimate of $\omega_\beta$. 
Whereas BD0+pTPSS tends to give a too short $R_e$ and too large 
frequencies, BD0+SCAN methods are, overall, 
in excellent agreement with CCSD(T). 
It is worth noting that the trends 
for the  UO$_2^{2+}$ frequencies are the same as those 
observed for a set of ten first- and second-row diatomics 
in Ref.~\cite{Garza2015}: 
CCD0+pTPSS overestimates the frequencies, while CCD0+SCAN 
combination improve results. 
The trend is similar for BD0+DFT methods, which are more accurate
than their CCD0+DFT counterparts. 

CCD0, BD0, and their combinations with DFT (in particular those 
with SCAN) fare well against other methods. 
Results from HF and KS-DFT methods in Table~\ref{tab:uo2} 
suggest that the description of UO$_2^{2+}$ is dependent 
on the amount of HF exchange in the functional: more HF 
exchange leads to shorter bond lengths and higher frequencies. 
This dependence can make common hybrids unreliable for 
high accuracy work. 
The ``cousin'' of CCD0, pCCD, underestimates 
$R_e$ more than all other methods except HF. 
In pCCD,  a singlet pairing scheme is also employed, 
but only the diagonal (optimized) terms are retained.
A better performance of CCD0 as compared to pCCD could be 
expected because the former contains more contributions 
in the cluster operator: the $T$ operator of pCCD is
\begin{equation}
  T_\text{pCCD} = \sum_{ia} t_i^a
  c_{a\uparrow}^\dag  c_{a\downarrow}^\dag
  c_{i\downarrow}  c_{i\uparrow}  , 
  \label{eq:TpCCD}
\end{equation}
which is only a part of the $T^{[0]}$ of CCD0. 
Although pCCD normally compensates for the missing terms via 
an orbital optimization, this optimization is nontrivial and 
can have multiple solutions.  
Nonetheless, pCCD has the advantage of having lower scaling. 
The cost of CCD0 is determined by the cost of solving the CCD0 
equations with symmetrized amplitudes. Thus, the scaling 
of CCD0 is the same as that of CCD, $\mathcal{O}(N^6)$. 
For pCCD, the cost of solving the pertinent CC equations is 
a remarkably low $\mathcal{O}(N^3)$, although a 
$\mathcal{O}(N^5)$ basis transformation transformation is required 
for the orbital optimization (and this is important for 
achieving good results and ensuring size consistency).
Compared to the traditional, ``gold-standard'' 
CCSD(T), CCD0 and CCD0+DFT are an order of magnitude lower 
in cost, while providing similar results and being more 
reliable for static correlation~\cite{Bulik2015}. 

Table~\ref{tab:uo2} also shows
CAS-srDFT results from Ref.~\cite{Fromager2009}. 
These methods
belong to a class of techniques that complement long-range 
wavefunction two-body energy with short-range DFT 
Hartree--exchange--correlation. 
The idea is to capture the dynamic correlation, which is short-range, 
with DFT and avoid double counting by range separation; 
an approach to avoid double counting that is very different 
from that used in CCD0+DFT. 
The CAS-srDFT bond lengths are comparable to those of 
CCD0+DFT, although the latter has the advantage that it does not 
neglect effects of static correlation in the short-range.
CASSCF and CASPT results are also comparable to those from 
CCD0/BD0 (and their +DFT combinations), whereas MP2
gives too large bond lengths.

\subsection{Neptunyl and Plutonyl (NpO$_\mathbf{2}^{\mathbf{3+}}$
and PuO$_\mathbf{2}^{\mathbf{4+}}$)}

The increased charge in neptunyl and plutonyl 
(NpO$_2^{3+}$ and PuO$_2^{4+}$) as compared to the 
isoelectronic uranyl enhances 
degeneracies, bolstering static correlation.
This makes these ions more challenging to describe than 
UO$_2^{2+}$. 
In fact, CCSD(T) results reported in the literature
for these systems come with a warning: 
The $T_1$ diagnostics for neptunyl and plutonyl are 
0.22 and 0.35, respectively~\cite{Straka2001}. 
Empirically, CCSD(T) predictions are considered 
unreliable when $T_1 > 0.2$~\cite{Lee1989} (although the norm 
of $T_2$ is probably a more reliable and better 
indicator of static correlation~\cite{Bulik2015}, 
we did not find these data in the literature).

\begin{table}
  \begin{center}
    \caption{Bond lengths ($R_e$ in \AA) and harmonic vibrational 
    frequencies ($\omega_e$ in cm$^{-1}$)
    for the neptunyl ion NpO$_2^{3+}$ ($D_{\infty h}$) calculated 
    by various methods.
    To the best of our knowledge, 
    no experimental data is available for this species.}
    \label{tab:npo2}
    \scalebox{0.90}{
      \begin{tabular*}{0.5\textwidth}{@{\extracolsep{\textwidth minus\textwidth}}  lcccc  }
        \hline
        This Work &    &                      &                     &                  \\
        Method & $R_e$ & $\omega_{\text{as}}$ & $\omega_{\text{s}}$ & $\omega_{\beta}$ \\
        \hline 
        CCD0        & 1.662 & 1183 & 1076 & 217 \\
        CCD0+pTPSS  & 1.645 & 1188 & 1121 & 221 \\
        CCD0+pSCAN  & 1.654 & 1157 & 1092 & 216 \\
        CCD0+prSCAN & 1.656 & 1150 & 1085 & 216 \\
        CCD0+trSCAN & 1.652 & 1161 & 1095 & 215 \\
        BD0         & 1.671 & 1142 & 1021 & 205 \\
        BD0+pTPSS   & 1.653 & 1135 & 1071 & 201 \\
        BD0+pSCAN   & 1.664 & 1095 & 1033 & 190 \\
        BD0+prSCAN  & 1.665 & 1091 & 1030 & 191 \\
        BD0+trSCAN  & 1.663 & 1095 & 1033 & 191 \\
        \hline
        Literature &    &                      &                     &                  \\
        Method & $R_e$ & $\omega_{\text{as}}$ & $\omega_{\text{s}}$ & $\omega_{\beta}$ \\
        \hline
        CCSD(T)$^a$       & 1.682 & 1106 &  990 & 141 \\
        MP2$^a$           & 1.757 & 900  &  879 & 106 \\
        CASSCF(12,12)$^a$ & 1.678 & -- & -- & -- \\
        CASSCF(12,16)$^a$ & 1.685 & -- & -- & -- \\

 \hline    

 \multicolumn{5}{@{}l}{\footnotesize{$^a$From Ref.~\cite{Straka2001}; 
     RSC+2$g$ on Np and aug-cc-pVDZ on O. }} \\
 \multicolumn{5}{@{}l}{\footnotesize{$^b$From Ref.~\cite{Fromager2009}; 
     RECP ($14s13p10d8f6g$)/[$6s6p5d4f3g$]
     on U and }} \\
 \multicolumn{5}{@{}l}{\footnotesize{($4s5p1d$)/[$2s3p1d$] on O. }} \\
      \end{tabular*}
    }
  \end{center}
\end{table}

Table~\ref{tab:npo2} shows $R_e$ and $\omega_e$ results for 
NpO$^{3+}_2$. 
The methods for which results are shown in this table all 
predict a linear geometry for neptunyl. 
The prediction of a linear geometry can be considered a success 
for CCD0+DFT methods because it has been shown that 
a correct description of both exchange and correlation is 
necessary for this~\cite{Fromager2009}. 
Other MR+DFT that avoid double counting 
typically do so at the cost of introducing some 
(inexact) DFT exchange. 
Popular KS-DFT functionals like LDA, PBE, and B3LYP predict
bent geometries for neptunyl~\cite{Fromager2009}. 
As a result, MR+DFT combinations having substantial DFT exchange 
also tend toward bent geometries~\cite{Fromager2009}. 
We also note that CCSD(T) gives results comparable to CASSCF, 
BD0, and BD0+DFT. 
Considering this and the 
fact that the $T_1$ diagnostic of CCSD(T)~\cite{Straka2001} 
($T_1 = 0.22$) is close to the limit 
of what is considered reliable~\cite{Lee1989} (0.20), 
it seems like the CCSD(T) results are salvageable for  
NpO$^{3+}_2$. 
MP2 is not reliable for the $f^0$ actinyl series 
isoelectronic to uranyl beyond uranyl itself; 
the bond lengths in neptunyl appear to be largely overestimated, 
while the frequencies are underestimated.

\begin{table}
  \begin{center}
    \caption{Bond lengths ($R_e$ in \AA) and harmonic vibrational 
    frequencies ($\omega_e$ in cm$^{-1}$)
    for the NUN molecule ($D_{\infty h}$) calculated 
    by various methods.}
    \label{tab:nun}
    \scalebox{0.90}{
      \begin{tabular*}{0.5\textwidth}{@{\extracolsep{\textwidth minus\textwidth}}  lcccc  }
        \hline
        This Work &    &                      &                     &                  \\
        Method & $R_e$ & $\omega_{\text{as}}$ & $\omega_{\text{s}}$ & $\omega_{\beta}$ \\
        \hline 
        CCD0        & 1.735 & 1098 & 1059 & 188 \\
        CCD0+pTPSS  & 1.717 & 1165 & 1102 & 119 \\
        CCD0+pSCAN  & 1.724 & 1140 & 1079 & 198 \\
        CCD0+prSCAN & 1.724 & 1151 & 1089 & 114  \\
        CCD0+trSCAN & 1.719 & 1162 & 1100 & 117 \\
        BD0         & 1.736 & 1109 & 1067 & 239 \\
        BD0+pTPSS   & 1.721 & 1142 & 1080 & 109 \\
        BD0+pSCAN   & 1.730 & 1111 & 1051 & 205 \\
        BD0+prSCAN  & 1.729 & 1129 & 1067 & 113 \\
        BD0+trSCAN  & 1.725 & 1136 & 1074 & 118 \\        
        CCSD        & 1.733 & 1119 & 1050 & 155 \\
        PBE         & 1.738 & 1085 & 1032 & $37i$ \\
        PBEh        & 1.713 & 1154 & 1113 & 118 \\
        LC-$\omega$PBE & 1.701 & 1200 & 1166 & 129 \\
        \hline
        Literature &    &                      &                     &                  \\
        Method & $R_e$ & $\omega_{\text{as}}$ & $\omega_{\text{s}}$ & $\omega_{\beta}$ \\
        \hline
        HF$^a$       & 1.640 & --- & --- & --- \\
        CCSD(T)$^a$ & 1.722 & --- & ---  & --- \\
        MP2$^a$ & 1.721 & --- & ---  & --- \\
        CAS-srLDA$^a$ & 1.710 & --- & ---  & --- \\
        CAS-srPBE$^a$ & 1.710 & --- & ---  & --- \\
        CASPT2(12,12)$^b$ & 1.735 & 1031  & 969  & --- \\
        Expt.$^c$             & --- & 1077 & --- & --- \\
 \hline    
 \multicolumn{5}{@{}l}{\footnotesize{$^b$From Ref.~\cite{Fromager2009}; 
     RECP ($14s13p10d8f6g$)/[$6s6p5d4f3g$]
     on U and }} \\
 \multicolumn{5}{@{}l}{\footnotesize{($4s5p1d$)/[$2s3p1d$] on O. }} \\
 \multicolumn{5}{@{}l}{\footnotesize{$^b$From Ref.~\cite{Gagliardi2000}; 
     RSC on U and $4s3p2d1f$ ANO-L on O. }} \\
 \multicolumn{5}{@{}l}{\footnotesize{$^c$From Ref.~\cite{Zhou1996}.}} \\
      \end{tabular*}
    }
  \end{center}
\end{table}

In the case of  PuO$_2^{4+}$, the CCD0 and BD0 geometry 
optimizations resulted in linear structures but with 
imaginary bending frequencies of 
149 and 227 cm$^{-1}$, respectively.
Attempts to optimize the structure starting from a bond angle 
of about 160$^\circ$ (a minimum at a fixed bond length) resulted 
in the geometry going back to linear with an extended bond 
length without achieving convergence after many iterations. 
Straka \textit{et al.}~\cite{Straka2001} reported 
that CAS(12,16) predicts the  PuO$_2^{4+}$ system to disintegrate
and that CCSD(T) results are highly unreliable 
($T_1$ = 0.35). 
Plutonium has four common oxidation states Pu(III), 
Pu(IV), Pu(V), and Pu(VI); the highest known oxidation 
state of Pu in aqueous solution is VII and is only stable in 
strong alkaline medium. 
However, 
the formal oxidation state of plutonium in PuO$_2^{4+}$
is VIII. 
The too-high charge of Pu in PuO$_2^{4+}$ brings
the actinide's low-lying $f$ orbitals closer to the atom and 
makes them less suitable for bonding. 
Considering all this, it seems very likely that PuO$_2^{4+}$ 
is not a stable gas phase species. 
Nevertheless, 
Tsushima~\cite{Tsushima} has theorized a possible synthesis for 
PuO$_5$OH$^{3-}$ in strong alkaline solution, and
Huang \textit{et al.}~\cite{Haung2015} have recently reported 
theoretical evidence for the stability of 
a PuO$_2$F$_4$ complex.

\subsection{Uranium Dinitride (NUN)}

The linear NUN molecule has been studied theoretically and 
and observed experimentally~\cite{Gagliardi2000,Zhou1996,Andrews2014}. 
The interest in this compound stems mostly from 
its similarity to the important uranyl ion, to which it is 
isoelectronic. 
Results for $R_e$ and $\omega_e$
computed by various methods here and in previous works
are listed on Table~\ref{tab:nun}. 
General trends are similar to those observed for uranyl: 
CCD0 and CCD0+DFT methods are in good agreement with 
available CASPT2 and CCSD(T) methods, although the CASPT2 
vibrational frequencies are somewhat lower than those predicted 
by coupled cluster methods. 
Results by KS-DFT methods depend on the amount of exchange,
with the bond length being reduced as more exchange is incorporated. 
In addition, PBEh and LC-$\omega$PBE yield a linear geometry but PBE 
shows an imaginary bending frequency for the linear structure. 
Again, HF gives bond lengths that are too-short, although MP2 
appears to give more reasonable bond lengths for this species.

\subsection{Thorium Oxides ThO and ThO$^\mathbf{2+}$} 

ThO is the most studied (experimentally and theoretically) of 
the actinide monoxides~\cite{Gagliardi2000}. 
ThO$^{2+}$ is also of interest because Th(IV) is the most 
common oxidation state for thorium. 
Interest in thorium chemistry arises mainly from potential 
applications of the thorium fuel cycle---the transmutation 
of the abundant $^{232}$Th into artificial $^{233}$U, which is 
the actual fuel in the nuclear chain reaction. 
In fact, thorium has been touted as potential 
``wonder fuel''~\cite{Ashley2012}
due to certain advantages over uranium like greater abundance 
and better resistance to nuclear weapons proliferation, 
albeit the latter advantage has been contended~\cite{Ashley2012}.

\begin{table}
  \begin{center}
    \caption{Bond lengths ($R_e$ in \AA) and harmonic vibrational 
    frequencies ($\omega_e$ in cm$^{-1}$)
    for ThO and ThO$^{2+}$ calculated 
    by various methods.}
    \label{tab:tho}
    \scalebox{0.90}{
      \begin{tabular*}{0.5\textwidth}{@{\extracolsep{\textwidth minus\textwidth}}  lcccc  }
        \hline
        & \multicolumn{2}{c}{ThO}  &  \multicolumn{2}{c}{ThO$^{2+}$}  \\
        \cline{2-3} \cline{4-5}
        Method & $R_e$ & $\omega_{\text{e}}$ & $R_e$ & $\omega_{\text{e}}$ \\
        \hline 
        CCD0        & 1.869 & 880 & 1.792 & 1000 \\
        CCD0+pTPSS  & 1.847 & 906 & 1.774 & 1030 \\
        CCD0+pSCAN  & 1.856 & 890 & 1.781 & 979 \\
        CCD0+prSCAN & 1.859 & 887 & 1.783 & 1009  \\
        CCD0+trSCAN & 1.853 & 893 & 1.778 & 1016 \\
        BD0         & 1.871 & 879 & 1.795 & 999 \\
        BD0+pTPSS   & 1.849 & 901 & 1.776 & 990 \\
        BD0+pSCAN   & 1.859 & 884 & 1.784 & 970 \\
        BD0+prSCAN  & 1.861 & 881 & 1.786 & 1000 \\
        BD0+trSCAN  & 1.856 & 886 & 1.781 & 1007 \\        
        CCSD        & 1.856 & 891 & 1.791 & 1008 \\
        PBE         & 1.841 & 896 & 1.776 & 1008 \\
        PBEh        & 1.828 & 928 & 1.760 & 1052 \\
        LC-$\omega$PBE & 1.821 & 946 & 1.752 & 1080 \\
        CASPT2$^a$  & 1.863 & 879 & 1.792 & 988 \\
        Expt.       & 1.840 & 895 & --- & --- \\
 \hline    
 \multicolumn{5}{@{}l}{\footnotesize{$^a$From 
     Refs.~\cite{Kovacs2015,Kovacs2011}; 
     using all-electron basis set. }} \\
 \multicolumn{5}{@{}l}{\footnotesize{$^b$From
     Refs.~\cite{Kovacs2015,Dewberry2007,Edvinson1984,Edvinson1965}}} \\
      \end{tabular*}
    }
  \end{center}
\end{table}

Table~\ref{tab:tho} compares the 
$R_e$ and $\omega_{\text{e}}$ for ThO and ThO$^{2+}$ obtained by various 
methods. The data from CCD0, BD0, and their combinations 
with DFT are all very close to available CASPT2 and experimental 
data from the literature. 
Once more, combinations using the SCAN 
functional tend to be closer to the reference CASPT2 values, though 
all the combinations presented here provide satisfactory results.
The trend of DFT methods to shorten the bond length as more Fock 
exchange is incorporated that was observed in the previous cases is also 
present here. 

\subsection{Nobelium Oxides NoO and NoO$_\mathbf{2}$}

The chemistry of Nobelium is largely uncharacterized, which offers 
a possibility for theoretical methods to provide unique insight into it. 
Table~\ref{tab:noo} shows bond lengths and harmonic
frequencies for NoO and NoO$_2$ computed by  BD0, BD0+DFT, and 
some standard coupled cluster and KS-DFT methods. 
The $\omega_e$ for NoO is omitted because the floppy bond 
of this molecule lead to somewhat large errors in the fitting 
procedure to obtain the force constants; however, they are 
consistently estimated to be around 550 cm$^{-1}$. 
Likewise, CCSD(T) data for NoO$_2$ is absent due to the 
difficulty in converging these calculations. 
To the best of our knowledge, there are no accurate reference data 
for these compounds in the literature.

\begin{table}
  \begin{center}
    \caption{Bond lengths ($R_e$ in \AA) and harmonic vibrational 
    frequencies ($\omega_e$ in cm$^{-1}$)
    for NoO and NoO$_2$ calculated 
    by various methods. For methods that predict geometries that deviate 
    from linearity in  NoO$_2$, the bond angle is shown in parenthesis. }
    \label{tab:noo}
    \scalebox{0.90}{
      \begin{tabular*}{0.5\textwidth}{@{\extracolsep{\textwidth minus\textwidth}}  lccccc  }
        \hline
        & NoO & \multicolumn{4}{c}{NoO$_2$} \\
        \cline{2-2} \cline{3-6}
        Method &  $R_e$ &  $R_e$ & $\omega_{\text{as}}$ & $\omega_{\text{s}}$ & $\omega_{\beta}$ \\
        \hline 
        BD0         & 2.008 & 1.868 & 703  & 583 & 256 \\
        BD0+pTPSS   & 1.948 & 1.828 & 665  & 628 & 193 \\
        BD0+pSCAN   & 1.972 & 1.844 & 636  & 600 & 183 \\
        BD0+prSCAN  & 1.979 & 1.849 & 666  & 630 & 181 \\
        BD0+trSCAN  & 1.966 & 1.840 & 682  & 645 & 184 \\
        CCSD        & 1.923 & 1.832 & 756  & 660 & 173 \\
        CCSD(T)     & 1.983 & ---   & ---  & --- & --- \\
        PBE         & 1.911 & 2.045(125$^\circ$) & 523 & 465 & 149 \\ 
        PBEh        & 1.903 & 2.036(129$^\circ$) & 516 & 467 & 142 \\
        LC-$\omega$PBE & 1.879 & 1.974(131$^\circ$) & 613 & 542 & 150 \\
 \hline    
      \end{tabular*}
    }
  \end{center}
\end{table}

The most noticeable feature of Table~\ref{tab:noo} is that 
BD0, BD0+DFT, and CCSD predict a linear geometry for NoO$_2$, 
whereas KS-DFT methods yield bent geometries. 
The bonds in the nobelium compounds are considerably larger 
than for the $f^0$ actinide oxides because the $5f$ orbitals 
are completely filled in the former. 
The effect of dynamic correlation is also much larger: 
addition of DFT correlation to BD0 leads to decrease in bond 
lengths of about 0.02--0.05 \AA, compared to changes the more modest 
changes (rarely more than 0.02 \AA) observed above. 
Based on the results for the previous actinide compounds, 
we could expect BD0+pSCAN and BD0+prSCAN to provide accurate 
estimates for the bond lengths and frequencies for the nobelium oxides. 
In the case of NoO, CCSD(T), BD0+pSCAN, and BD0+prSCAN results 
are highly similar.

\section{Conclusions and Outlook}

Singlet-paired coupled cluster and its combinations with 
DFT can provide accurate estimates for properties of actinide 
compounds such as geometries and vibrational frequencies. 
Compared to the data from the most accurate estimates
available (CASPT, CCSD(T), experiment), typical deviations 
of CCD0 and CCD0+DFT methods are around 0.01 \AA \, for
bond lengths and 20 cm$^{-1}$ for harmonic vibrational 
frequencies. 
These deviations are similar to the ones computed 
for simple first- and second-row diatomics in a previous 
work~\cite{Garza2015}, indicating the wide applicability 
and generality of the approach. 
The CCD0 and CCD0+DFT results presented here reinforce 
predictions by other methods (\textit{e.g.}, CCSD(T) or CASSCF) for 
species for which no experimental data are available 
(\textit{e.g.}, the important UO$_2^{2+}$ cation),
including the instability of the PuO$_2^{4+}$ ion. 
The BD0 and BD0+DFT results for NoO and NoO$_2$ are probably 
the most reliable estimates available so far, as previous reports 
(see Ref.~\cite{Kovacs2015})
utilized KS-DFT functionals which, 
according also to the results here, are not consistently 
reliable for actinide compounds. 

For most of the molecules studied here, CCSD(T) 
appears to provide reliable results. CCD0+DFT 
can provide data of similar quality to CCSD(T) 
while being a order in magnitude lower in cost and 
much more robust for handling static correlation. 
The CCD0 and CCD0+DFT methods are seen to be more 
accurate than pCCD for UO$_2^{2+}$.
Because pCCD has lower cost than CCD0 and 
misses mostly dynamic correlation, the results here 
suggest that pCCD+DFT combinations analogous to CCD0+DFT 
(such as those suggested in 
Refs.~\cite{Garza2015b,Garza2015,Garza2015c}) may be 
very promising for actinide chemistry. 
Currently, we are working on some of these pCCD+DFT combinations 
and on extensions of CCD0 for treating open-shells, which would 
greatly increase the applicability of CCD0-based techniques to 
actinide chemistry.

\section*{Acknowledgments}

This work was supported by the U.S.
Department of Energy, Office of Science, Office of Basic Energy
Sciences, Heavy Element Chemistry Program under Award
Number DE-FG02-04ER15523. G.E.S. is a Welch Foundation
Chair (C-0036).


\end{document}